\documentclass[prb, twocolumn,amsfonts,amsmath,amssymb,eufrak]{revtex4-1}

\usepackage{epsfig}
\usepackage{color}
\usepackage{bm}

\usepackage{braket}

\usepackage{graphicx}
\usepackage{amsthm}

\usepackage{multirow}

\newcommand{\eq}[1]{\begin{equation} #1 \end{equation}}
\newcommand{\eqa}[2]{\begin{equation} #1 \label{#2} \end{equation}}

\newcommand{\mx}[1]{\begin{pmatrix}#1 \end{pmatrix}}

\newcommand{\figin}[4]
{\begin{figure}[tb]
\centering
\includegraphics[width= #1]{#2.pdf}
\caption{#3}
\label{f:#4}
\end{figure}}

\newcommand{\todayd}{\the\year/\the\month/\the\day}

\newcommand{\bib}{\bibitem}

\newcommand{\lb}{\label}

\newcommand{\Tr}{\mathrm{Tr}}

\newcommand{\bel}{\begin{easylist}}
\newcommand{\eel}{\end{easylist}}
\newcommand{\bi}[1]{\begin{itemize} #1 \end{itemize}}
\newcommand{\be}[1]{\begin{enumerate} #1 \end{enumerate}}

\newcommand{\eref}[1]{Eq.~\eqref{#1}}
\newcommand{\tref}[1]{Theorem.~\ref{t:#1}}
\newcommand{\dref}[1]{Definition.~\ref{t:#1}}

\newcommand{\pref}[1]{Proposition.~\ref{t:#1}}
\newcommand{\fref}[1]{Fig.~\ref{f:#1}}
\newcommand{\sref}[1]{Sec.~\ref{s:#1}}

\def \({\left(}
\def \){\right)}
\def \[{\left[}
\def \]{\right]}

\newcommand{\abs}[1]{\left|#1\right|}

\newcommand{\sumtwo}[2]%
{\mathop{\sum_{#1}}_{#2}}
\newcommand{\sumthree}[3]%
{\mathop{\mathop{\sum_{#1}}_{#2}}_{#3}}
\newcommand{\sumfour}[4]%
{\mathop{\mathop{\mathop{\sum_{#1}}_{#2}}_{#3}}_{#4}} 
\newcommand{\prodtwo}[2]%
{\mathop{\prod_{#1}}_{#2}}
\newcommand{\mintwo}[2]%
{\mathop{\min_{#1}}_{#2}}
\newcommand{\maxtwo}[2]%
{\mathop{\max_{#1}}_{#2}}
\newcommand{\maxthree}[3]%
{\mathop{\mathop{\max_{#1}}_{#2}}_{#3}}
\newcommand{\limtwo}[2]%
{\mathop{\lim_{#1}}_{#2}}
\newcommand{\suptwo}[2]%
{\mathop{\sup_{#1}}_{#2}}
\newcommand{\supthree}[3]%
{\mathop{\mathop{\sup_{#1}}_{#2}}_{#3}}
\newcommand{\supfour}[4]%
{\mathop{\mathop{\mathop{\sup_{#1}}_{#2}}_{#3}}_{#4}} 
\newcommand{\inftwo}[2]%
{\mathop{\inf_{#1}}_{#2}}
\newcommand{\infthree}[3]%
{\mathop{\mathop{\inf_{#1}}_{#2}}_{#3}}
\newcommand{\inffour}[4]%
{\mathop{\mathop{\mathop{\inf_{#1}}_{#2}}_{#3}}_{#4}} 

\newcommand\calP{{\cal P}}

\newcommand{\bse}{\boldsymbol{e}}

\newcommand{\bsi}{\boldsymbol{i}}
\newcommand{\bsj}{\boldsymbol{j}}

\newcommand{\bsv}{\boldsymbol{v}}

\newcommand{\bsx}{\boldsymbol{x}}

\newcommand{\bszero}{\boldsymbol{0}}

\newcommand{\bbC}{\mathbb{C}}

\newcommand{\para}[1]{\medskip {\bf #1}}

\newtheorem{thm}{Theorem}
\newtheorem{lm}[thm]{Lemma}
\newtheorem{pro}[thm]{Proposition}
\newtheorem{co}[thm]{Corollary}

\newcommand{\bpf}[1]{\begin{proof} #1 \end{proof}}

\newcommand{\bthm}[1]{\begin{thm} #1 \end{thm}}

\newcommand{\bpro}[1]{\begin{pro} #1 \end{pro}}
\newcommand{\bco}[1]{\begin{co} #1 \end{co}}

\theoremstyle{definition}
\newtheorem{dfn}{Definition}
\newcommand{\bdf}[1]{\begin{dfn} #1 \end{dfn}}

\def\rnum#1{\resizebox{0.5em}{\height}{\expandafter{\romannumeral #1}}}
\def\Rnum#1{\resizebox{0.5em}{\height}{\uppercase\expandafter{\romannumeral #1}}}
\newcommand{\ha}{h^{(1)}}
\newcommand{\hb}{h^{(2)}}

\makeatletter
\renewcommand{\@cite}[1]{\textsuperscript{#1)}}
\makeatother

\begin{document}

\preprint{APS/123-QED}

\newcommand{\nameoftitle}{
Frustration-freeness is decidable but unprovable
}

\title{
\begin{flushleft}
{\huge \bf \nameoftitle}
\end{flushleft}
}

\author{{\large Naoto Shiraishi}}
\affiliation{Graduate school of arts and sciences, The University of Tokyo, 3-8-1 Komaba, Meguro-ku, Tokyo 153-8902, Japan}%



\maketitle




\onecolumngrid

\vspace{-10pt}

\noindent
{\bf
The frustration-free property is valuable in quantum many-body systems, which states that its ground states simultaneously minimize all local interaction terms, even when these terms act on overlapping degrees of freedom.
Prominent examples include the AKLT model and Kitaev's toric code model, which serve as paradigmatic models of symmetry-protected topological phases and topological order, respectively.
A fundamental question is whether a given Hamiltonian is frustration-free for all system sizes.
Here we establish two seemingly contradictory results concerning this decision problem. 
First, we prove that frustration-freeness is decidable for any fixed local Hilbert space dimension. 
Specifically, there exists a finite, Hamiltonian-independent system-size cutoff such that frustration-freeness at this cutoff size guarantees frustration-freeness at all system sizes. 
Second, on the other hand, we also show that there exists a frustration-free Hamiltonian whose frustration-freeness cannot be proved within the standard axioms of mathematics (ZFC).
We reconcile these seemingly contradictory results by showing that the cutoff, although finite for each local Hilbert space dimension, is uncomputable as a function of that dimension. 
Our findings reveal an unexpected fundamental distinction between decidability and provability arising in quantum many-body physics.
}

\bigskip

\twocolumngrid

Frustration-free Hamiltonians serve as valuable theoretical stages for studying the low-energy physics of strongly correlated quantum many-body systems~\cite{Tasbook, Cir21}. 
A Hamiltonian is frustration-free if its ground states simultaneously minimize all local interaction terms, even when these terms overlap and do not commute. 
This property plays a central role in the theory of tensor-network states, where matrix-product states and projected entangled-pair states can be realized as exact ground states of frustration-free Hamiltonians~\cite{Gar07}. 
In addition to these benefits, frustration-freeness imposes strong constraints on low-energy physics. 
In frustration-free gapless systems, the finite-size energy gap obeys inverse-square-type bounds under broad conditions~\cite{LL25, MSW24, MSW25}. 
Frustration-free systems with commuting-projector Hamiltonians cannot realize integer and fractional Hall effect with nonzero Hall conductance~\cite{KF20, KS20, ZLB22}.
These results demonstrate that frustration-freeness not only facilitates the analytical study of quantum many-body systems but also constrains their possible low-energy properties.

A fundamental question is whether a given set of local Hamiltonians gives rise to frustration-free Hamiltonians for all system sizes. 
Necessary and sufficient conditions for frustration-freeness have been derived for certain restricted classes of models, including free-fermion systems~\cite{Ono26} and generic one-dimensional qudit chains~\cite{Mov10}. 
More general sufficient conditions have been obtained through the quantum Lov\`{a}sz local lemma~\cite{Sat16, He19}, which relates frustration-freeness to the ranks and overlap structure of local constraints. 
Building on the quantum 2-SAT algorithm~\cite{Bra06, Ara15}, efficient methods have also been developed to determine frustration-freeness in finite spin-1/2 systems with two-body interactions~\cite{Bea10}.

The perspective of theoretical computer science have provided powerful tools for understanding fundamental limitations of quantum many-body physics.
Determining whether a system has a spectral gap~\cite{CGW15} or whether it thermalizes~\cite{SM21} has been shown to be undecidable in suitably general settings. 
In the former case, the undecidability still arises in shift-invariant nearest-neighbor interacting systems on two-dimensional square lattices with a fixed local Hilbert space dimension.
We remark that the fate of decidability can change drastically when the class of Hamiltonians is restricted.
For example, the spectral-gap problem is decidable for translation-invariant, nearest-neighbor frustration-free spin-1/2 chains in one dimension~\cite{BG15}. 
These contrasting results raise a fundamental question: whether frustration-freeness is decidable when the local Hilbert space dimension is fixed?

Here we investigate the frustration-free problem for translation-invariant, nearest-neighbor Hamiltonians on two-dimensional square lattices from the perspective of computability theory and mathematical logic. 
We establish two seemingly contradictory results. 
\be{
\item[A.] For any fixed local Hilbert space dimension $d$, there exists a finite system-size cutoff $L(d)$, independent of the Hamiltonian, such that frustration-freeness at this cutoff size guarantees frustration-freeness at all system sizes. 
Hence, frustration-freeness is decidable for each fixed $d$. 
\item[B.] Assuming the consistency of the standard axioms of mathematics (ZFC), we show that there exists a Hamiltonian which is frustration-free for all system sizes, yet whose frustration-freeness cannot be proved within ZFC. 
}
We reconcile these seemingly contradictory results by proving that no algorithm computes the cutoff $L(d)$ for all $d$. 
Our findings reveal a striking distinction between the existence of a finite decision criterion and our ability to compute and justify it, arising from the interplay between Hilbert's basis theorem and G\"{o}del's second incompleteness theorem.

\figin{8cm}{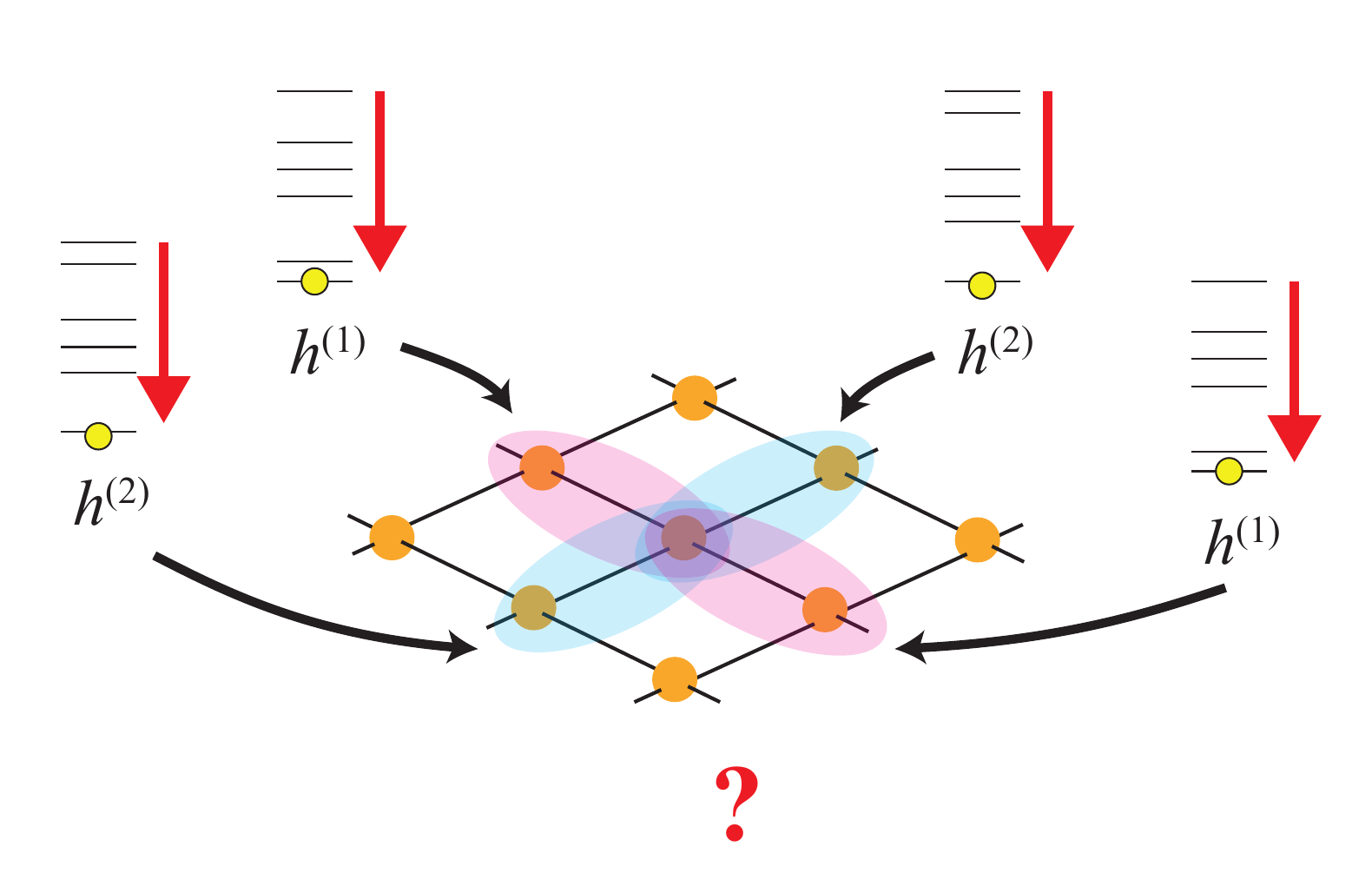}{
A frustration-free Hamiltonian admits a ground state minimizing all the local Hamiltonians simultaneously.
In the frustration-free problem on two-dimensional square lattices, given a set of local Hamiltonians we decide whether the corresponding shift-invariant system is frustration-free for all system sizes.
\tref{cutoff} states that there exists a finite cutoff $L(d)$ which depends only on the local Hilbert space dimension $d$ such that the frustration-freeness on the $L(d)\times L(d)$ square lattice guarantees the frustration-freeness at all system sizes.
On the other hand, \tref{unprovable} states that if the standard axioms of mathematics (ZFC) is consistent, then there exists a set of local Hamiltonians which is frustration-free for all system sizes while we cannot prove its frustration-freeness within ZFC.
}{FF-schematic}

\para{Frustration-free problem}

In this article, we consider a quantum spin systems on an $L\times L$ two-dimensional square lattice with shift-invariant nearest-neighbor interaction with the open boundary condition.
We denote the dimension of the local Hilbert space (i.e., the size of a single spin) by $d$.
The Hamiltonian is given by
\eqa{
H=\sum_{i_1=1}^{L-1}\sum_{i_2=1}^L \ha_{\bsi, \bsi+\bse_1}+\sum_{i_1=1}^{L}\sum_{i_2=1}^{L-1}\hb_{\bsi, \bsi+\bse_2},
}{H}
where $\bsi:=(i_1,i_2)$ is the position on the lattice, and we denote by $\bse_1:=(1,0)$ and $\bse_2:=(0,1)$.
The Hamiltonian can be anisotropic, i.e., $\ha\neq \hb$.
Shifting the origin of the energy if needed, without loss of generality, we suppose that $\ha$ and $\hb$ are positive semi-definite and its lowest energy is zero.

The subject of this article is the frustration-free property.
We say that a Hamiltonian with a system size $L$ is frustration-free if there exists a state $\ket{\Psi}$ such that $\ha_{\bsi, \bsi+\bse_1}\ket{\Psi}=0$ and $\hb_{\bsj, \bsj+\bse_2}\ket{\Psi}=0$ for all legal $\bsi$ and $\bsj$ (\fref{FF-schematic}).
Namely, in the frustration-free system we can minimize all the interaction Hamiltonians simultaneously, even though the system has complicated interactions and thus in general we face conflicts between local Hamiltonians.
We also say that a Hamiltonian is frustration-free without further qualification if this Hamiltonian is frustration-free for all system sizes $L$.

A fundamental question is whether a given set of local interactions yields frustration-free Hamiltonians for all system sizes. 
For any fixed system size, frustration-freeness can be determined by finite-dimensional linear algebra. 
Thus the challenge is to determine whether this property holds for all system sizes.

We formulate this question as a decision problem. 
A decision problem is called {\it decidable} if there exists a Turing machine that halts on every valid input and correctly outputs either YES or NO. 
Otherwise, the problem is called {\it undecidable}.
In our problem setting, for a fixed local Hilbert space dimension $d$, the input consists of the local interaction matrices, and the task is to determine whether the resulting Hamiltonian is frustration-free for all system sizes. 
Here, we implicitly restrict the matrix elements to algebraic numbers so that each input can be specified by a finite sequence of bits.

\para{Decidability of frustration-free property}

We find that for a fixed local Hilbert space dimension $d$, the decision problem of the frustration-free property is decidable.
More precisely, there exists a finite cutoff size $L(d)$ such that if a system is frustration-free at $L=L(d)$, then this system is also frustration-free for all system sizes.

\bthm{[Existence of finite cutoff]\lb{t:cutoff}
For fixed local Hilbert space dimension $d$, there exists a finite cutoff $L(d)$ such that if a given Hamiltonian in the form of \eref{H} is frustration-free for system size $L(d)\times L(d)$, then this Hamiltonian is also frustration-free for any system size $L$.
}

\bco{[Decidability]\lb{t:decidable}
For fixed local Hilbert space dimension $d$, whether a given Hamiltonian \eqref{H} is frustration-free for all system sizes is decidable.
}

Although we formulate the decidability theorem only for two-dimensional square lattices, the same argument applies more generally to any fixed sequence of finite graphs, with a fixed finite set of local interactions assigned to their edges. 
We emphasize that the space of possible inputs is infinitely large even with fixing the local Hilbert space dimension $d$.
Indeed, the spectral-gap problem is undecidable for the same input space with a fixed $d$~\cite{CGW15}, ensuring that our input space is sufficiently rich to support undecidable decision problems.
This richness comes from the fact that the matrix elements of local Hamiltonians can encode arbitrarily long descriptions. 
Thus, fixing the local Hilbert space dimension does not, by itself, guarantee decidability.

At first sight, the existence of a finite cutoff seems to provide a straightforward way to verify frustration-freeness: it suffices to examine a single finite system at the cutoff size. 
However, as the next theorem reveals, the existence of a finite cutoff does not guarantee that frustration-freeness is provable.

\para{Unprovable frustration-free system}

To discuss provability, we adopt {\it Zermelo–Fraenkel set theory with the axiom of choice} (ZFC), which is a standard axiomatic foundation of modern mathematics~\cite{Kunbook}. 
Most of modern mathematics, including algebra, analysis, and topology, can be formalized within ZFC. 
We therefore use provability within ZFC as a formal notion of provability in standard mathematics.

We say that a theory is {\it inconsistent} if this theory proves both a statement and its negation.
Otherwise, a theory is called {\it consistent}.
As is widely believed, we assume that ZFC is consistent.

\bthm{[Unprovability]\lb{t:unprovable}
Suppose the consistency of ZFC.
Then, there exists a Hamiltonian in the form of \eref{H} such that (i) this Hamiltonian is frustration-free, but (ii) we cannot prove that this Hamiltonian is frustration-free within ZFC.
}

This theorem reveals a fundamental limitation that the frustration-freeness of a Hamiltonian may be true yet unprovable within our standard mathematics. 
Importantly, this is not merely an abstract existence result, but we can explicitly construct this Hamiltonian.
Although this Hamiltonian is so simple that it has only commuting local interactions, its frustration-freeness cannot be proved within ZFC.

This phenomenon is not specific to ZFC but extends to a broad class of axiomatic systems. 
More precisely, let $T$ be any consistent, computably axiomatizable theory to which G\"{o}del's second incompleteness theorem applies. 
Then there exists a frustration-free Hamiltonian whose frustration-freeness cannot be proved within $T$.

\figin{8cm}{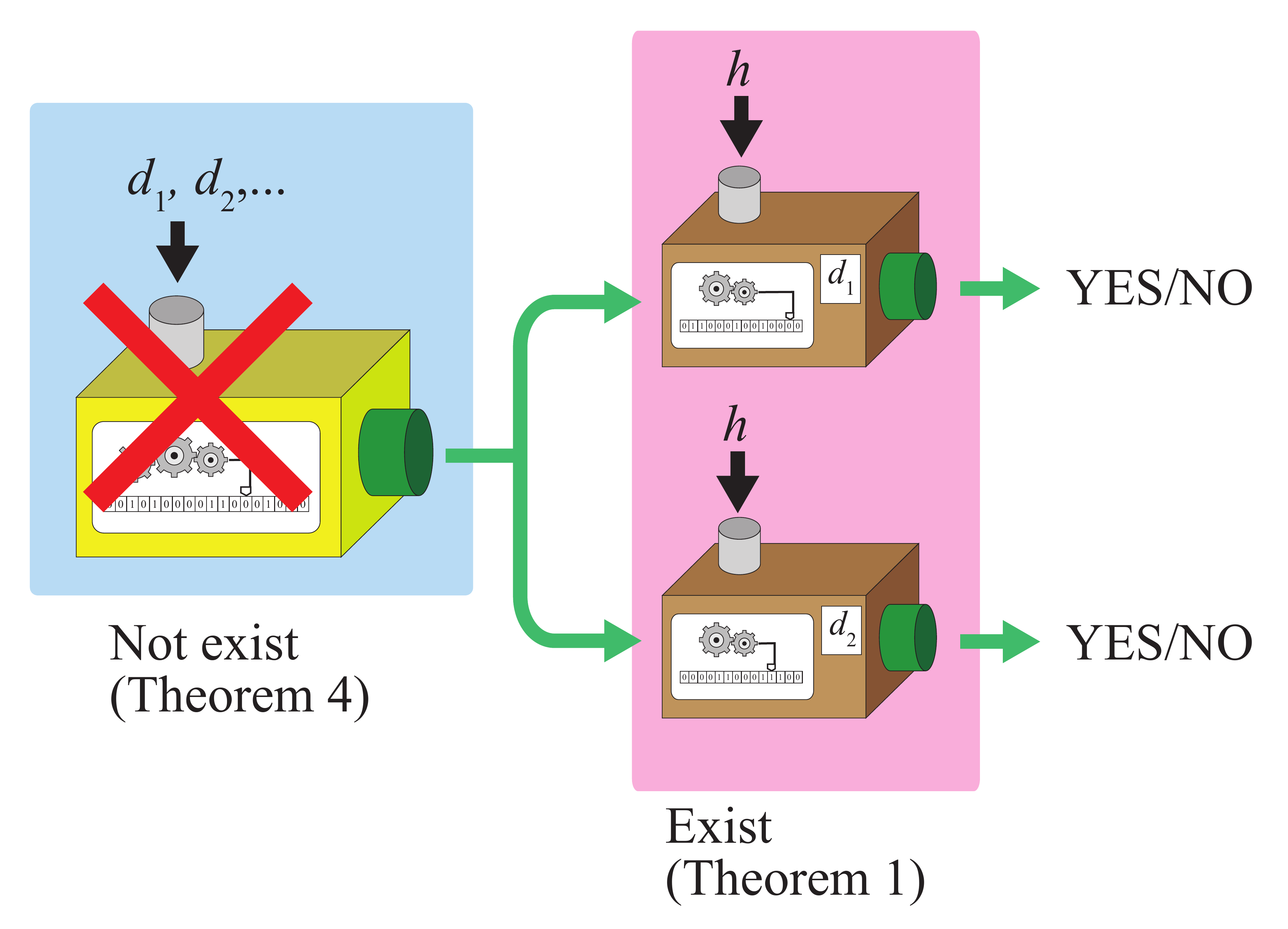}{
For each local Hilbert space dimension $d$, there exists a Turing machine deciding whether a given set of local Hamiltonians is frustration-free for all system sizes.
However, importantly this machine cannot be constructed uniformly in $d$.
Namely, there is no algorithm constructing the above machines uniformly from $d$.
}{nonuniform}

\para{Reconciling decidability and unprovability}

The decidability result (\tref{cutoff} and Corollary~\ref{t:decidable}) appears to contradict the unprovability result (\tref{unprovable}). 
Indeed, the existence of a finite cutoff suggests that frustration-freeness can be established by examining only a finite system. 
If the cutoff could be explicitly identified and justified, such a finite calculation would provide a proof of frustration-freeness within ZFC. 
The key to reconciling these results is the distinction between the existence of a finite cutoff and the ability to determine it. 
In fact, we show that the cutoff $L(d)$ is uncomputable as a function of $d$.

\bthm{[Uncomputability of the cutoff]\lb{t:uncomputable-cutoff}
No algorithm can compute a valid cutoff $L(d)$ in \tref{cutoff} for every local Hilbert space dimension $d$.
}

This uncomputability of $L(d)$ follows from the undecidability of the frustration-free problem when the local Hilbert space dimension $d$ is not fixed, which can be established using a construction similar to that in \tref{unprovable}.
If $L(d)$ were computable, we could decide frustration-freeness for arbitrary $d$ by first computing $L(d)$ and then testing a single finite system, contradicting the undecidability. 
In fact, we can also show that the cutoff grows at least as fast as the busy-beaver function $\mathrm{BB}(n)$, in the sense that $L(an+b)\geq\mathrm{BB}(n)$ for some constants $a$ and $b$. 
Consequently, the decidability established in \tref{cutoff} is non-uniform in $d$: although a decision algorithm exists for each fixed $d$, no algorithm can construct such decision algorithms uniformly from $d$ (\fref{nonuniform}).

The crucial point is that decidability requires only the existence of a Turing machine that correctly decides the problem, not an explicit construction of such a machine. 
In many familiar decidability results, a decision algorithm is constructed explicitly, and therefore this distinction does not matter.
In the present case, however, the distinction is essential. 
Despite the existence of a finite cutoff for every fixed $d$, no algorithm can compute a valid cutoff uniformly in $d$. 
More remarkably, let $d^*$ be the local Hilbert space dimension of the unprovable frustration-free Hamiltonian in \tref{unprovable}. 
Although ZFC proves the existence of a finite cutoff for this fixed dimension $d^*$, it cannot prove that any particular integer is a valid cutoff.
The existence of a decision algorithm does not necessarily imply that its correctness can be established within ZFC.
This illustrates a fundamental distinction between the existence of a decision algorithm and our ability to establish its correctness within ZFC.

From a mathematical perspective, these results reveal a subtle interplay between Hilbert's basis theorem~\cite{AMbook} and G\"{o}del's second incompleteness theorem~\cite{Smibook}. 
Hilbert's basis theorem guarantees that infinitely many polynomial constraints can be reduced to finitely many constraints, but does not provide a general procedure for identifying a suitable finite subset. 
Although algebraic techniques such as Gr\"{o}bner bases frequently enable explicit computations in many algebraic problems~\cite{Ben17, Sei71, Hon00, FS01, Sch21}, no such uniform procedure is possible in the frustration-free problem. 
Thus, frustration-freeness provides a concrete physical setting where the existence of a finite decision criterion and the ability to determine or justify that criterion are fundamentally distinct.

\para{Proof outline of decidability and unprovability}

Theorem~\ref{t:cutoff} follows from Hilbert's basis theorem applied to polynomial constraints expressing the existence of a common zero-energy state. 
Fixing a basis of the local Hilbert space, we represent two local Hamiltonians $\ha$ and $\hb$ as $d^2\times d^2$ matrices. 
Their matrix elements thus form a $2d^4$-tuple of variables, whose number is independent of the system size $L$. 
Using basic techniques in linear algebra, we can rewrite the condition of the frustration-freeness at size $L$ in terms of the common zeros of a finite set of polynomials $P_L$.

Now consider the infinite set of polynomials $\calP=\bigcup_{L=2}^\infty P_L$. 
The common zeros of $\calP$, restricted to admissible local Hamiltonians, are precisely the frustration-free Hamiltonians for all system sizes. 
Hilbert's basis theorem guarantees that the common zeros of $\calP$ are completely determined by its finite subset $\tilde{\calP}\subseteq\calP$. 
Since $\tilde{\calP}$ contains only a finite number of constraints, there exists a finite $L(d)$ such that $\tilde{\calP}\subseteq\bigcup_{L=2}^{L(d)}P_L$. 
Hence, frustration-freeness for $L\leq L(d)$ guarantees frustration-freeness for any system size. 
Furthermore, because the frustration-freeness at size $L(d)$ implies the frustration-freeness at size $L'$ less than $L(d)$, it suffices to check frustration-freeness at the single size $L(d)$. 
This establishes the finite-cutoff theorem.

\medskip

\tref{unprovable} follows by combining the theory of Wang tilings with G\"{o}del's second incompleteness theorem. 
For any Turing machine $M$, one can construct a finite set of Wang tiles that can tile the infinite plane if and only if $M$ does not halt~\cite{Rob71}. 
We then construct a corresponding quantum system and its Hamiltonian by assigning each tile to a local basis state and introducing local interaction Hamiltonians that penalize forbidden pairs of neighboring tiles. 
Specifically, these terms are projection operators onto the subspaces spanned by forbidden pairs. 
The resulting Hamiltonian is frustration-free for all system sizes if and only if the tiles can cover the infinite plane, which is equivalent to the situation that $M$ does not halt.

Now consider a Turing machine that systematically searches for a contradiction in ZFC and halts if and only if ZFC is inconsistent~\cite{YA16}. 
Encoding this machine into a Hamiltonian through the above Wang tiling construction yields a Hamiltonian whose frustration-freeness is equivalent to the consistency of ZFC. 
Note that this equivalence can itself be established within ZFC. 
Assuming that ZFC is consistent, the Hamiltonian is therefore frustration-free. 
However, if ZFC could prove its frustration-freeness, it could also prove its own consistency, contradicting G\"{o}del's second incompleteness theorem. 
Thus, the frustration-freeness of this explicitly constructed Hamiltonian is unprovable within ZFC.

\para{Discussion}

Our results reveal a fundamental distinction between mathematical decidability and the ability to construct and justify a decision algorithm. 
Importantly, fixing the local Hilbert space dimension $d$ does not make the space of possible inputs finite: local Hamiltonians with finite exact descriptions indeed form an infinite set with unbounded description lengths. 
Indeed, the spectral-gap problem is undecidable for the class of two-dimensional, translation-invariant, nearest-neighbor Hamiltonians at certain fixed $d$~\cite{CGW15}. 
Thus, the decidability of frustration-freeness is not a trivial consequence of fixing the local Hilbert space dimension.


From a mathematical perspective, our approach is related to the Hilbert method~\cite{Boj19}, which exploits Hilbert's basis theorem to establish the existence of finite sets of constraints. 
In several previous applications of the Hilbert method and related techniques~\cite{Ben17, Sei71, Hon00, FS01, Sch21, May84}, additional algebraic structures have enabled the construction of explicit decision algorithms. 
In contrast, our result establishes decidability without providing a uniform algorithm for determining the required cutoff. 
Indeed, for certain fixed local Hilbert space dimensions, even a concrete upper bound on the cutoff cannot be established within ZFC. 

In some familiar examples, nonuniform decidability follows from the fact that possible inputs can be reduce to a finite set~\cite{MNS93, Del00, CDT06, CDF97}. 
In other settings, finite-size criteria may depend on individual inputs, without providing a common cutoff applicable to all inputs~\cite{Bau74}. 
A notable exception was established by Aminof {\it et al.}~\cite{Ami14} in the parameterized model checking problem, where a common cutoff exists for fixed logical parameters, while unfixed parameters can render the problem undecidable. 
However, their example of such nonuniformity relies on specially constructed network topologies, which the authors themselves describe as pathological. 
In contrast, the frustration-free problem has a genuinely infinite input space even for fixed $d$, sufficiently rich to support undecidable decision problems, as demonstrated by the spectral-gap problem. 
Nevertheless, we establish nonuniform decidability with a Hamiltonian-independent finite cutoff on the ordinary two-dimensional square lattice. 
Thus, our result provides an unusual example of nonuniform decidability with a common cutoff, arising naturally in a problem of quantum many-body physics.

\bigskip

{\bf \large Method}

{

\para{Frustration-free problem}

We consider the Hamiltonian $H_L$ defined in Eq.~\eqref{H} on an $L\times L$ square lattice with open boundary conditions, where the same local interactions $h^{(1)}$ and $h^{(2)}$ are used for all system sizes $L$. 
Without loss of generality, we assume that these local interactions are positive semidefinite and have zero ground-state energy. 
For a fixed $L$, the Hamiltonian $H_L$ is called frustration-free if there exists a state $\ket{\Psi_L}$ satisfying $h^{(a)}_{\bsi,\bsi+\bse_a}\ket{\Psi_L}=0$ for any local interaction term ($a=1,2$). 
Since all local terms are positive semidefinite, this condition is equivalent to the existence of a zero-energy ground state of $H_L$. 
When we say that the Hamiltonian is frustration-free without specifying the system size, we mean that $H_L$ is frustration-free for every $L\geq 2$.

An important property of these open-boundary Hamiltonians is that frustration-freeness at size $L$ implies frustration-freeness at any smaller size $L'\leq L$. 
To see this, let $\ket{\Psi_L}$ be a frustration-free ground state of $H_L$, and consider its reduced density matrix $\rho_{L'}$ on an $L'\times L'$ sublattice. 
Since all local interaction within this sublattice annihilate $\ket{\Psi_L}$, we have $\Tr[h^{(a)}_{\bsi,\bsi+\bse_a}\rho_{L'}]=0$ for all bonds of the smaller lattice. 
Due to the positive semidefiniteness of the local Hamiltonians, every eigenvector of $\rho_{L'}$ with a nonzero eigenvalue is annihilated by all these local terms, implying that any such eigenvector provides a frustration-free ground state of $H_{L'}$.

For a fixed pair of local interactions, define a binary sequence $\{a_L\}_{L\geq 2}$ by setting $a_L=0$ if $H_L$ is frustration-free and $a_L=1$ otherwise. 
The above property implies that once a term in this sequence takes 1, then all other terms after this term should take 1. 
Therefore, it takes one of two possible forms: 
\bi{
\item $a_L=0$ for all $L$ (Frustration-free case).
\item There exists a first non-frustration-free system size $L_c$ such that $a_L=0$ for $2\leq L<L_c$ and $a_L=1$ for all $L\geq L_c$ (Non-frustration-free case).
}
The frustration-free problem amounts to deciding which of these two possibilities occurs. 
Importantly, although a non-frustration-free Hamiltonian can be identified by finding a finite system size with $a_L=1$, checking increasingly large systems does not by itself establish frustration-freeness for all sizes.
We also remark that the monotonicity of $\{a_L\}$ does not imply the existence of a finite common upper bound of $L_c$ for all Hamiltonians.

\para{Frustration-free problem as common zeros}

We now show that frustration-freeness can be expressed as common zeros of a finite set of polynomials in the matrix elements of the local Hamiltonians. 
Fixing a basis of the local Hilbert space, we represent the two local Hamiltonians $h^{(1)}$ and $h^{(2)}$ as $d^2\times d^2$ matrices. 
Their matrix elements $h^{(a)}_{pq,rs}$ ($a=1,2$ and $1\leq p,q,r,s\leq d$) can be regarded as $2d^4$ variables, whose number is independent of the system size $L$. 
As required before, the local Hamiltonians are assumed to be positive-semidefinite and have zero ground energies.

For a given system size $L$, let $N_L=d^{L^2}$ be the dimension of the full Hilbert space and $B_L=2L(L-1)$ be the number of nearest-neighbor bonds. 
We denote by $h_{e_j}$ ($j=1,\ldots,B_L$) the local interaction operator acting on the $j$-th bond, embedded in the full Hilbert space as a  tensor product of identity operators on all other sites. 
Frustration-freeness is equivalent to the existence of a nonzero vector $\bsv\in\mathbb C^{N_L}$ satisfying $h_{e_j}\bsv=\bszero$ for all $j$. 
To express these conditions collectively, we introduce the $B_LN_L\times N_L$ matrix
\eq{
M_L=\begin{pmatrix}
h_{e_1}\\
h_{e_2}\\
\vdots\\
h_{e_{B_L}}
\end{pmatrix}.
}
By construction, the system is frustration-free if and only if there exists a nonzero vector $\bsv$ satisfying $M_L\bsv=\bszero$, which is equivalent to $\operatorname{rank}M_L<N_L$.
This rank condition is equivalent to the vanishing of all determinants of $N_L\times N_L$ submatrices of $M_L$. 
Let $M_L[I]$ denote the square submatrix formed by selecting $N_L$ rows of $M_L$, where $I=\{k_1,\ldots,k_{N_L}\}\subseteq\{1,\ldots,B_LN_L\}$. 
Then, frustration-freeness is equivalent to
\eq{
\det M_L[I]=0
}
for all such choices of $I$.

Crucially, any matrix element of $M_L$ is linear in the $2d^4$ variables $h^{(a)}_{pq,rs}$, and hence each determinant $\det M_L[I]$ is a polynomial in these variables. 
We denote by $P_L$ the finite set of polynomials obtained from all possible choices of $I$. 
For physically admissible local Hamiltonians, frustration-freeness on the $L\times L$ square lattice is equivalent to the simultaneous vanishing (i.e., common zero) of all polynomials in $P_L$.
Here, the physical requirements that the local Hamiltonians are Hermitian, positive semidefinite, and have zero ground-state energies are imposed separately and need not be incorporated into $P_L$.

Importantly, although the number and degrees of these polynomials grow with $L$, the number of variables remains fixed at $2d^4$. 
Therefore, the polynomial constraints for all system sizes belong to the same polynomial ring; $\bbC[x_1,\ldots,x_{2d^4}]$. 
This observation allows us to apply Hilbert's basis theorem to the infinitely many constraints associated with all system sizes.

\para{Noetherianity and Hilbert's basis theorem}

Let $R$ be a commutative ring and $T\subseteq R$.
We define the ideal generated by $T$ as
\eq{
I[T]:=\left\{\sum_{i\in J}a_ix_i\;\middle|\;a_i\in R,\ x_i\in T,\ J\text{ is a finite set}\right\}.
}
A ring is called {\it Noetherian} if every ideal is finitely generated~\cite{AM69}.
Equivalently, for any subset $T\subseteq R$, there exists a finite subset $S\subseteq T$ such that $I[T]=I[S]$. 
Hilbert's basis theorem states that a polynomial ring in finitely many variables over a Noetherian ring is also Noetherian. 
Since the complex field $\bbC$ is Noetherian, the polynomial ring $\bbC[x_1,\ldots,x_m]$ is also Noetherian for any finite $m$.

An important consequence of Noetherianity is that the common zeros of infinitely many polynomials can be characterized by only a finite number of polynomials. 
To see this, consider a set of polynomials $T\subseteq\mathbb C[x_1,\ldots,x_m]$ and define its common zero set as
\eq{
Z(T):=\{ \bsx\in \bbC^{\otimes m}| f(\bsx)=0 \text{ for all } f\in T\}.
}
Since every polynomial in $I[T]$ vanishes at all common zeros of $T$, we have $Z(T)=Z(I[T])$. 
By Noetherianity, there exists a finite subset $S\subseteq T$ satisfying $I[T]=I[S]$, which means
\eq{
Z(T)=Z(I[T])=Z(I[S])=Z(S).
}
Thus, in terms of the common zeros an infinite collection of polynomial equations in a fixed number of variables is equivalent to its suitable finite subcollection.

We now apply this result to establish the existence of a finite cutoff for the frustration-free problem. 
Let $\calP=\bigcup_{L=2}^{\infty}P_L$ be the union of the polynomial constraints obtained in the preceding section for all system sizes. 
A pair of physically admissible local Hamiltonians $(h^{(1)},h^{(2)})$ is frustration-free for all system sizes if and only if its matrix elements belong to $Z(\calP)$. 
Since all polynomials in $\calP$ belong to the Noetherian ring $\bbC[x_1,\ldots,x_{2d^4}]$, there exists a finite subset $\widetilde{\calP}\subseteq\calP$ such that
\eq{
Z(\calP)=Z(\widetilde{\calP}).
}
As $\widetilde{\calP}$ contains only finitely many polynomials, there exists a finite system size $L^*$ satisfying
\eq{
\widetilde{\calP}\subseteq\bigcup_{L=2}^{L^*}P_L.
}
Consequently,
\eq{
Z(\calP)=Z\left(\bigcup_{L=2}^{L^*}P_L\right),
}
which means that frustration-freeness for all $L\leq L^*$ guarantees frustration-freeness for any system size. 
Furthermore, since frustration-freeness on an $L\times L$ square lattice with open boundary conditions implies frustration-freeness on any smaller square lattice as explained before, it suffices to examine the single system size $L^*$. 
This establishes the existence of a finite cutoff $L(d)$ that is independent of the specific local Hamiltonians.

\para{Wang tiles and Turing machines}

We first introduce the Wang tiling problem. 
Consider a finite set of square tiles $A^1,\ldots,A^n$, each with a color assigned to each of its four edges. 
The tiles can be placed on a square lattice without rotation, and two neighboring tiles are legal only if the colors of their shared edges match. 
In an abstract manner, we can specify the legal pairs of neighboring tiles separately for the horizontal and vertical directions. 
The Wang tiling problem asks whether these tiles can cover the entire infinite two-dimensional plane without violating any adjacency constraints.

A fundamental result in computability theory is that Wang tiles can simulate all possible computations of Turing machines~\cite{Rob71}. 
In a suitable construction, the states of a Turing machine and the symbols on its tape are encoded in tile configurations, and the local matching rules enforce the machine's transition rules. 
Roughly speaking, horizontal rows of tiles represent spacial configurations of the machine, and its time evolution is encoded in the vertical direction. 
Importantly, for any given Turing machine $M$, we can construct a finite set of tiles which tiles the infinite plane if and only if $M$ does not halt.
Note that some additional techniques ensure this equivalence without prescribing an initial tile or a boundary condition.

We next connect Wang tilings to frustration-free Hamiltonians. 
For a given set of $n$ tiles, we construct a quantum spin system with local Hilbert space dimension $d=n$, whose orthonormal basis states $\ket{A^1},\ldots,\ket{A^n}$ represent the possible tiles. 
We define the local interaction in direction $a=1,2$ as
\eq{
h^{(a)}=\sum_{(p,q)\in F_a}\ket{A^pA^q}\bra{A^pA^q},
}
where $F_a$ denotes the set of illegal pairs of neighboring tiles in direction $a$. 
Each local Hamiltonian is thus a projection operator that assigns zero energy to legal pairs and unit energy to illegal pairs. 
Hence, the Hamiltonian is frustration-free if and only if the infinite two-dimensional plane admits a valid tiling.
Notably, all local Hamiltonians in this construction are diagonal and mutually commuting.

In conclusion, we obtain an effective correspondence between any Turing machine $M$ and a Hamiltonian that is frustration-free if and only if $M$ does not halt.

\para{Decidability and its nonuniformity}

We clarify an important distinction between decidability and uniform decidability. 
A decision problem is decidable if there exists a Turing machine that halts on every valid input and correctly outputs YES or NO. 
Notably, this definition requires only the existence of such a machine, and decidability does no necessarily provide its explicit construction.

Now consider a family of decision problems indexed by a parameter $n$. 
We say that the family is {\it pointwise decidable} if, for every fixed $n$, there exists a Turing machine $M_n$ that correctly decides the problem for every input $x$. 
However, this does not necessarily imply {\it uniform decidability}, which requires a single Turing machine $\widetilde{M}$ that takes both $n$ and $x$ as inputs and correctly decides the problem for any pair $(n,x)$, i.e., $\widetilde{M}(n,x)=M_n(x)$. 
In other words, the existence of individual decision algorithms $M_n$ does not guarantee a computable procedure that constructs such an algorithm from $n$.

In our frustration-free problem, the parameter $n$ corresponds to the local Hilbert space dimension $d$. 
The finite-cutoff theorem establishes pointwise decidability for every fixed $d$, whereas the undecidability of the problem with unrestricted $d$ rules out uniform decidability.

\para{Scaling of $L(d)$}

We further investigate how rapidly the cutoff $L(d)$ grows with the local Hilbert space dimension $d$. 
To this end, we first introduce the busy-beaver function ${\rm BB}(m)$, which is defined as the maximum number of 1s left on the tape by any halting Turing machine with $m$ internal states and a binary tape alphabet starting from a blank tape. 
Since we should have a tape longer than or equal to $k$ in order to leave $k$ copies of 1s on the tape, there exists a halting $m$-state Turing machine that requires at least ${\rm BB}(m)$ tape cells to complete its computation without reaching a tape boundary.
It is known that the busy-beaver function is uncomputable and grows faster than any computable function.
Indeed, the value of ${\rm BB}(5)$ was computed at 2024~\cite{BB5}, and the value of ${\rm BB}(6)$ is unknown at present, which confirms the hardness of computation of the busy-beaver function even for very small $n$.

We now relate the busy-beaver function to the cutoff using the Wang-tiling construction described above. 
For a fixed tape alphabet, an $m$-state Turing machine can be encoded into a set of Wang tiles whose number grows at most linearly with $m$ \cite{DL25, Karnote}. 
Moreover, the construction can be arranged so that a halting Turing machine requiring $k$ tape cells gives rise to a tile set that cannot tile the infinite plane but admits valid tilings of all finite squares with side lengths below $k$. 
The corresponding Hamiltonian is therefore frustration-free up to these system sizes but not for all system sizes. 
By absorbing the fixed overheads of the construction into the constants, we obtain
\eq{
L(am+b)\geq BB(m)
}
for some positive constants $a$ and $b$. 
Thus, the finite cutoff grows at least as rapidly as the busy-beaver function, up to a linear rescaling of its argument.

\para{AI declarations}

NS used ChatGPT (GPT-5.4, GPT-5.5 Pro, GPT-5.6 Sol, and GPT-6 Astra) to establish the dicidability result and clarify how to understand this result and the unprovable result consistently.
NS also used ChatGPT (GPT-6 Astra) to polish the English presentation.

}
\bigskip

{\bf \normalsize Date Availability}

Data sharing not applicable to this article as no datasets were generated or analysed during the current study.

\bigskip

{\bf \normalsize Acknowledgement}

NS thanks Mizuki Sanatani for fruitful discussion.
This work was supported by Grant-in-Aid for Early-Career Scientists 26K17048, Grant-in-Aid for Transformative Research Areas (B) 26K00021, and JST ERATO Grant Number JPMJER2302.

\bigskip

{\bf \normalsize Author contributions}

NS has responsibility to all the results.

\bigskip

{\bf \normalsize Competing Interest}

The author declares no competing interest.

\let\oldaddcontentsline\addcontentsline
\renewcommand{\addcontentsline}[3]{}

\let\addcontentsline\oldaddcontentsline

\newpage

\onecolumngrid


\newcommand{\vo}{\upsilon}
\newcommand{\midskip}{\vspace{3pt}}

\setcounter{thm}{0}
\renewcommand{\thethm}{S.\arabic{thm}}
\renewcommand{\thedfn}{S.\arabic{dfn}}
\renewcommand{\thelm}{S.\arabic{lm}}
\renewcommand{\thepro}{S.\arabic{pro}}
\setcounter{figure}{0}
\renewcommand{\thefigure}{S.\arabic{figure}}

\setcounter{equation}{0}

\makeatletter
\def\shorttableofcontents#1#2{\bgroup\c@tocdepth=#2\@restonecolfalse
  \settowidth\js@tocl@width{\headfont\prechaptername\postchaptername}%
  \settowidth\@tempdima{\headfont\appendixname}%
  \ifdim\js@tocl@width<\@tempdima \setlength\js@tocl@width{\@tempdima}\fi
  \ifdim\js@tocl@width<5zw \divide\js@tocl@width by 5 \advance\js@tocl@width 4zw\fi
\if@tightshtoc
    \parsep\z@
  \fi
  \if@twocolumn\@restonecoltrue\onecolumn\fi
  \@ifundefined{chapter}%
  {\section*{{#1}
        \@mkboth{\uppercase{#1}}{\uppercase{#1}}}}%
  {\chapter*{{#1}
        \@mkboth{\uppercase{#1}}{\uppercase{#1}}}}%
  \@startshorttoc{toc}\if@restonecol\twocolumn\fi\egroup}
\makeatother


\newpage

\setcounter{page}{1}

\begin{center}
{\large \bf Supplemental Material for  \protect \\ 
  ``\nameoftitle'' }\\
\vspace*{0.3cm}
Naoto Shiraishi \\
\vspace*{0.1cm}

{Graduate school of arts and sciences, The University of Tokyo} 

\end{center}

\tableofcontents

\section{Decidability}
\subsection{Review of commutative algebra}
We here review basic results of commutative algebra and Noetherian rings.
Throughout this section, we assume that a ring $R$ is a commutative ring otherwise explicitly noted.

\subsubsection{Ideals and Noetherian rings}

Consider a commutative ring $R$.
An ideal $I$ of a ring $R$ is a subset of $R$ satisfying the following properties.

\bdf{[Ideal]
A subset $I$ of a commutative ring $R$ is an {\it ideal} if
\bi{
\item For any $x,y\in I$, we have $x+y\in I$.
\item For any $x\in I$ and $a\in R$, we have $ax\in I$.
}
}

If the ring $R$ is not commutative, the above definition serves as the definition of the {\it left ideal}.

Any ideal $I$ contains the additive identity, $0\in I$, since the second condition asserts $0x=0\in I$.
Any ring has two ideals, the {\it unit ideal} $I=R$ and the {\it zero ideal} $I=\{0\}$, both of which are called {\it trivial ideals}.
An ideal which is not a trivial ideal is called a proper ideal.

We here present a basic but important fact that a field $F$ has only trivial ideals.
This fact is shown as follows.
Let $I$ be an ideal which is not the zero ideal.
Let $a\in I$ be a nonzero element of $I$.
Then, setting $x=a^{-1}$, we have $aa^{-1}=1\in I$.
Now, by setting $x=1$, any element $a\in R$ is also in $I$, since $1a=a\in I$.
Hence, $I$ is the unit ideal $I=R$.

We next introduce the generation of ideals.

\bdf{[Ideal generation]
The ideal generated by a set $S$ is defined as
\eq{
I[S]=\{ \sum_{i\in J}a_ix_i|a_i\in R, x_i\in S, J \text{ is a finite set}\}.
}
If an ideal $I$ can be obtained as an ideal generated by a finite set, we say that $I$ is a {\it finitely generated ideal}.
}

Now we are ready to introduce the Noetherian ring.

\bdf{[Noetherian ring]
A ring $R$ is a Noetherian ring if all ideals of $R$ are finitely generated.
}

Equivalently, a ring $R$ is Noetherian if, for any subset $T\subseteq R$, there exists a finite subset $S\subseteq T$ such that $I[T]=I[S]$. 
To see why $S$ can be chosen as a subset of $T$, recall that Noetherianity guarantees that the ideal $I[T]$ is generated by finitely many elements $S'=\{y_1,\ldots,y_k\}\subseteq I[T]$. 
By the definition of $I[T]$, each $y_i$ can be expressed as a finite linear combination of elements of $T$ with coefficients in $R$:
\eq{
y_i=\sum_{j=1}^{m_i}c_{ij}x_{ij},\qquad x_{ij}\in T,\quad c_{ij}\in R.
}
Hence, the finite set $S=\{x_{ij}\}_{i,j}\subseteq T$ satisfies $I[S]=I[T]$, as desired.

Since a field $F$ has only trivial ideals, any field $F$ is obviously a Noetherian ring.
This fact is used in the proofs of our main theorems.

\subsubsection{Hilbert's basis theorem and common zeros}

Let $R[X_1,\ldots, X_m]$ be the ring of polynomials with variables $X_1,\ldots , X_m$ over $R$.
The Hilbert's basis theorem states that the Noetherianity is inherited to its polynomial ring.

\bthm{[Hilbert's basis theorem]
If $R$ is a Noetherian ring, then for any finite $m$, $R[X_1,\ldots, X_m]$ is also a Noetherian ring.
}

Applying Hilbert's basis theorem to ideals of a polynomial ring, we find an important result on common zeros of polynomials.

\bdf{[Common zeros of a set of polynomials]
Consider a commutative Noetherian ring $R$.
Let $T \in R[X_1,\ldots , X_m]$ be a set of polynomials with $m$ variables.
We define its common zeros as
\eq{
Z(T):=\{ \bsx\in R^{\otimes m}| f(x)=0 \text{ for all } f\in T\}.
}
}

\bpro{\lb{t:common-zero}
Consider a commutative Noetherian ring $R$.
Let $T \in R[X_1,\ldots , X_m]$ be an infinite set of polynomials with $m$ variables.
Then, there exists a finite subset $S\subset T$ such that
\eq{
Z(T)=Z(S).
}
}

This proposition states that the common zeros of infinitely many polynomials can be completely characterized by a finite subset of these polynomials. 
This reduction from infinitely many constraints to finitely many plays a key role in establishing the existence of a finite cutoff.

\bpf{[Proof of \pref{common-zero}]
Let $I[T]$ be the ideal of $ R[X_1,\ldots , X_m]$ generated by a set of polynomials $T=\{ f_a\}_{a=1}^\infty$:
\eq{
I[T]=\{ \sum_{i\in 1}^n c_i f_{a_i}| n<\infty, c_i \in R\}.
}
Since every polynomial in $I[T]$ vanishes at all common zeros of $T$, we have $Z(T)=Z(I[T])$.

Hilbert's basis theorem states that $I[T]$ is a Noetherian ring.
Namely, there exists a finite set of polynomials $S=\{ g_1,\ldots , g_k\}\in I[T]$ which also generate $I[T]$, i.e., $I[T]=I[S]$.
This directly means
\eq{
Z(T)=Z(I[T])=Z(I[S])=Z(S).
}
}

In the next section, we use this proposition in the case of the complex field; $R=\bbC$.

\subsection{Frustration-free problem as common zeros}
\subsubsection{Problem setting}

We here generalize the stage of the frustration-free problem from a two-dimensional square lattice to a general graph sequence.
Let $d$ be the dimension of the local Hilbert space on a single site.
We prepare a set of $m$ local Hamiltonians $h^{1},\ldots , h^{m}$ on two sites.
Changing the origin of the local energy if necessary, without loss of generality we require that these local Hamiltonians are positive semi-definite and its ground energy is zero.
We fix a basis of the single-site Hilbert space and express these local Hamiltonians as $d^2\times d^2$ matrices whose $(ab,cd)$ matrix element is $h^{\alpha}_{ab,cd}=\bra{a}\bra{b}h^{\alpha}\ket{c}\ket{d}$.
To specify these matrix elements by a finite bit sequence, we restrict possible matrix elements to algebraic numbers.

We consider an infinite sequence of computable graphs $\{G_n\}_{n=1}^\infty$. 
Let $E_n$ and $V_n$ be sets of (directed) edges and vertices of $G_n$, respectively.
We define a function $f_n$ from an edge $e\in E_n$ to $\{ 1,\ldots ,m\}$, by which we assign a local Hamiltonian on this edge.
The Hamiltonian on graph $G_n$ is then given by
\eqa{
H(G_n)=\sum_{e\in E_n} h^{f_n(e)}_e.
}{H-G}

If a Hamiltonian $H(G_n)$ has a zero energy eigenstate, this state minimizes all local Hamiltonians on this graph $G_n$.
The frustration-free problem asks whether these Hamiltonians $H(G_n)$ have zero energy eigenstates for all $n$.


\bdf{[Graph-sequence frustration-free problem]\lb{t:FFproblem-gen}
Fix the dimension of local Hilbert space of a single site as $d$, an infinite sequence of computable graphs $\{G_n\}_{n=1}^\infty$, and a function $f_n$ from an edge $e\in E_n$ to $\{ 1,\ldots ,m\}$, with which we obtain a sequence of Hamiltonians $\{H(G_n)\}_{n=1}^\infty$ by \eref{H-G}.

Given a set of local Hamiltonians $\{h^\alpha\}_{\alpha=1}^m$ acting on two sites, decide whether $H(G_n)$ is frustration-free for all $n$.
}

The graph-sequence frustration-free problem includes the two-dimensional square-lattice model considered in the main text as a special case. 
This general framework accommodates lattices of arbitrary spatial dimensions and geometries, including honeycomb and kagome lattices, as well as aperiodic and disordered graph structures. 
Neither translational symmetry nor any particular geometric structure is required. 
Moreover, the graph sequence need not be nested, i.e., $G_n$ need not be a subgraph of $G_{n+1}$.
This point is revisited in \sref{obc}.

The only assumptions are that the local Hilbert space dimension $d$ and the number of distinct local interactions $m$ are fixed, and that the graph sequence $\{G_n\}$ and the assignment of interaction types to edges are specified by computable rules. 
The local interactions themselves, which constitute the input of the decision problem, are assumed to be algebraic numbers in order to admit finite exact descriptions. 
Under these assumptions, the finite-cutoff theorem applies to any such graph sequence, with a cutoff that may depend on the graph sequence and the interaction-assignment rule but is independent of the specific local Hamiltonians.

We note in passing that although computability of the graph sequence is assumed for an effective specification of the problem, the existence of a finite cutoff itself in fact does not rely on this assumption.
A related point is discussed in \sref{cutoff-remark}.

\subsubsection{Existence of finite cutoff}

We prove the decidability of the graph-sequence frustration-free problem.

\bthm{\lb{t:FF-cutoff-gen}
In the setting of \dref{FFproblem-gen}, there exists a finite cutoff $n(d)$ which depends on the dimension $d$, the graph sequence $\{G_n\}_{n=1}^\infty$ and function $f$, but does not depend on a set of local Hamiltonians $\{h^\alpha\}_{\alpha=1}^m$ such that if $H(G_{n})$ is frustration-free for all $n\leq n(d)$, then $H(G_n)$ is frustration-free for all $n$.

As its corollary, the graph-sequence frustration free-problem given in \dref{FFproblem-gen} is decidable.
}

\bpf{
To prove this, we reformulate the frustration-free problem as a problem of common zeros.
To this end, we fix a basis of the local Hilbert space, where the local interactions $h^{\alpha}$ are represented as $d^2\times d^2$ matrices. 
Their matrix elements $h^{\alpha}_{ab,cd}$ ($1\leq \alpha\leq m$ and $1\leq a,b,c,ds\leq d$) can thus be regarded as $md^4$ variables, whose number is independent of the system size $L$. 
As required, the local Hamiltonians are assumed to be positive-semidefinite and have zero ground energies.

The Hamiltonian $H(G_n)$ is frustration-free if and only if there exists a nonzero $d^{\abs{V_n}}$-dimensional vector $\bsv$ such that 
\eq{
h_e \bsv=\bszero
}
for all $e$.
Here, $h_e$ is a $d^{|V_n|}\times d^{|V_n|}$ matrix acting on the full Hilbert space, given by $h_e=h^{f_n(e)}\otimes\bigotimes_{i\in V_n\setminus e}I_i$, up to a permutation of tensor factors, where $I_i$ is the identity operator on vertex $i$.
Using $h_e$'s, we introduce a $\abs{E_n}d^{\abs{V_n}}\times d^{\abs{V_n}}$ matrix $M_n$ defined by
\eq{
M_n=\mx{h_{e_1} \\ h_{e_2} \\ \vdots \\ h_{e_{\abs{E_n}}}}.
}
By construction, $M_n$ satisfies
\eq{
M_n\bsv=\bszero,
}
which means that $d^{\abs{V_n}}$ column vectors of $M_n$ are linearly dependent.
This is equivalent to the rank condition 
\eqa{
{\rm rank}M_n\leq d^{\abs{V_n}}-1.
}{M-rank-cond}

Now we rewrite the rank condition \eqref{M-rank-cond} as a set of polynomial equations.
Let $M_n[k]$ be the $k$-th row vector of $M_n$.
Since the row rank and column rank of a matrix are equal, the rank condition \eqref{M-rank-cond} is equivalent to the condition that any set of $d^{\abs{V_n}}$ row vectors of $M_n$ is linearly dependent. 
Equivalently, the determinant of any square submatrix formed by selecting $d^{\abs{V_n}}$ distinct rows of $M_n$ must vanish:
\eqa{
\det \mx{M_n[k_1] \\ M_n[k_2] \\ \vdots \\ M_n[k_{d^{\abs{V_n}}}]}=0
}{M-det-cond}
for all choices of distinct row indices $k_1,\ldots,k_{d^{\abs{V_n}}}$.
Importantly, every matrix element of $M_n$ is 0, 1, or one of the $md^4$ variables $h^{\alpha}_{ab,cd}$, and hence \eref{M-det-cond} is a polynomial equation of $md^4$ variables $h^{\alpha}_{ab,cd}$ ($1\leq \alpha\leq m$, $1\leq a,b,c,d\leq d$).
There are at most $\mx{\abs{E_n}d^{\abs{V_n}} \\ d^{\abs{V_n}}}$ polynomial equations corresponding to the all possible choices of $k_1,\ldots , k_{d^{\abs{V_n}}}$.
Let $P_n$ be a set of all of these polynomial equations.
For physically admissible local Hamiltonians, frustration-freeness on the graph $G_n$ with function $f_n$ is equivalent to the simultaneous vanishing (i.e., common zero) of all polynomials in $P_n$.
Here, we put the physical requirements that the local Hamiltonians are Hermitian, positive semidefinite, and have zero ground-state energies, which are imposed separately from $P_n$.

We introduce the union of the polynomial constraints for all $n$ denoted by
\eq{
\calP=\bigcup_{n=1}^\infty P_n.
}
A given set of physically admissible local Hamiltonians $\{h^\alpha\}_{\alpha=1}^m$ is frustratoin-free, i.e., $H(G_n)$ is frustration-free for all $n$, if and only if this set is a common zero of $\calP$ (i.e., $Z(\calP)$).
The corollary of the Hilbert's basis theorem (\pref{common-zero}) states that there exists a finite subset $\widetilde{\calP}\subseteq\calP$ such that
\eq{
Z(\calP)=Z(\widetilde{\calP}).
}
As $\widetilde{\calP}$ contains only finitely many polynomials, there exists a finite system size $n^*$ satisfying
\eq{
\widetilde{\calP}\subseteq\bigcup_{n=2}^{n^*}P_n.
}
This means that frustration-freeness for all $n\leq n^*$ guarantees frustration-freeness for any $n$. 
This $n^*$ is what we desired, $n(d)$.

Testing whether finite number of polynomials are zero is obviously decidable.
}

\subsection{Remarks}


\subsubsection{Boundary conditions and cutoffs on square lattices}\lb{s:obc}

The graph-sequence frustration-free problem introduced in \dref{FFproblem-gen} encompasses a wide class of systems, including those with periodic boundary conditions. 
In particular, \tref{FF-cutoff-gen} does not require the graph sequence to be nested, i.e., $G_n$ need not be a subgraph of $G_{n+1}$. 
Without this nesting property, however, frustration-freeness at a given single system size does not necessarily imply frustration-freeness at smaller sizes. 
Consequently, the general finite-cutoff theorem requires checking $H(G_n)$ for all $n\leq n(d)$, rather than only at $n=n(d)$.

In contrast, consider a nested sequence of graphs satisfying the following two conditions:
\be{
\item[(i)] $G_n$ is a subgraph of $G_{n+1}$
\item[(ii)] The assignments of local interactions are consistent, $f_{n+1}|_{E(G_n)}=f_n$. 
}
Under these conditions, frustration-freeness of $H(G_n)$ implies frustration-freeness of $H(G_m)$ for all $m\leq n$. 
In fact, a frustration-free ground state on $G_n$, when restricted to $G_m$, yields a reduced density matrix supported on the common kernel of all local Hamiltonians on $G_m$. 
Therefore, to establish frustration-freeness for all system sizes, it suffices to check the single Hamiltonian $H(G_{n(d)})$. 
This applies, in particular, to $D$-dimensional lattices with open boundary conditions, as \tref{cutoff} in the main article.

The distinction is relevant to periodic boundary conditions, for which frustration-freeness need not be monotonic in the system size. 
For example, consider spin-1/2 systems with $h=\ket{00}\bra{00}+\ket{11}\bra{11}$. 
This Hamiltonian leads to the two degenerate ground states in the form of $\ket{010101\cdots}$.
On a chain with length $L$ with periodic boundary conditions, this Hamiltonian is frustration-free for even $L$ but not for odd $L$, since the alternating order is not consistently elongated at the boundary for odd $L$. 
Thus, checking frustration-freeness at a single system size does not generally suffice to confirm the frustration-freeness for all system size.

To summarize this distinction, we state the finite-cutoff theorems for open and periodic boundary conditions separately. 
We denote the corresponding Hamiltonians by $H_L^{\mathrm{OBC}}$ and $H_L^{\mathrm{PBC}}$, respectively, where the latter is obtained by adding the interactions across the periodic boundaries to Eq.~\eqref{H}.

\bthm{[Finite cutoff for frustration-freeness with open boundary conditions (same as \tref{cutoff})]\lb{t:cutoff-s}
For every fixed local Hilbert space dimension $d$, there exists a finite cutoff $L_{\mathrm{OBC}}(d)$, independent of the local interactions, such that if $H_{L}^{\mathrm{OBC}}$ is frustration-free for $L=L_{\mathrm{OBC}}(d)$, then $H_L^{\mathrm{OBC}}$ is frustration-free for any $L$.
}

\bthm{[Finite cutoff for frustration-freeness with periodic boundary conditions]\lb{t:cutoff-pbc}
For every fixed local Hilbert space dimension $d$, there exists a finite cutoff $L_{\mathrm{PBC}}(d)$, independent of the local interactions, such that if $H_L^{\mathrm{PBC}}$ is frustration-free for all $L\leq L_{\mathrm{PBC}}(d)$, then $H_L^{\mathrm{PBC}}$ is frustration-free for any $L$.
}

Thus, frustration-freeness is decidable for every fixed local Hilbert space dimension under both open and periodic boundary conditions, although the finite-size criteria required for these two cases are different.

\subsubsection{Comments on the cutoff $n(d)$}\lb{s:cutoff-remark}

We here puts two remarks on the cutoff $n(d)$.

First, the existence of a finite cutoff does not require the graph sequence $\{G_n\}$ to be computable or nested. 
In fact, even for an arbitrary noncomputable sequence of finite graphs, Noetherianity guarantees that the frustration-free constraints for the entire sequence can be characterized by those from a finite initial segment before a cutoff. 
At first sight, it may seem counterintuitive that examining only finitely many graphs suffices to determine frustration-freeness for an infinite sequence with no computable structure. 
The key fact to understand this counterintuitive situation is that the cutoff is allowed to depend on the entire graph sequence, including graphs appearing arbitrarily far along it. 
Thus, the cutoff is not an a priori bound independent of the graph sequence. 
In fact, no cutoff depending only on $d$ can work uniformly for all graph sequences, since new constraints can be introduced at arbitrarily large indices.
The existence of a finite cutoff is established by identifying a finite initial segment whose constraints already capture those of the entire infinite sequence after the cutoff.

Second, our uncomputability theorem (\tref{uncomputable-cutoff}) rules out a uniform algorithm for determining the cutoff in the two-dimensional square-lattice problem, but does not preclude the determination of explicit cutoffs for particular graph sequences and small local Hilbert space dimensions $d$. 
Nevertheless, even for a simple graph sequence, identifying a valid cutoff or a rigorous upper bound is not necessarily straightforward. 
This situation is reminiscent of the busy-beaver problem, where determining exact values can be extremely challenging even for Turing machines with very few internal states. 
For example, the busy-beaver value for five-state Turing machines was determined only in 2024~\cite{BB5}, while its six-state value remains unknown. 
Determining explicit cutoffs for small $d$ in physically relevant graph families is therefore an uneasy but interesting direction for future research.

\section{Unprovability}

\subsection{Review of theoretical computer science}

\subsubsection{Halting problem and imcompleteness theorem}\lb{s:review-halting}

We first review the notion of decidability, which plays a central role in our theorems.

\bdf{[Decidable and undecidable]
A decision problem is {\it decidable} if there exists a Turing machine that halts on every valid input after finitely many steps and correctly outputs YES or NO. 
If no such Turing machine exists, the problem is {\it undecidable}.
}

We make three remarks on this definition. 

First, the input to a Turing machine must be specified by a finite binary string. 
In our setting, the inputs are local Hamiltonians, whose matrix elements are therefore restricted to algebraic numbers. 
Any algebraic number can be specified exactly by a finite binary string, for example, by an integer polynomial together with information identifying one of its roots. 
Moreover, arithmetic operations and equality tests on algebraic numbers can be performed algorithmically. 
We emphasize that the restriction to algebraic matrix elements is needed to formulate the problem as a Turing-machine decision problem, whereas the existence of a finite cutoff holds for arbitrary matrix elements.

Second, decidability requires the Turing machine to halt on every valid input, regardless of whether the correct answer is YES or NO. 
A procedure that halts and outputs YES for every YES instance but may run forever on NO instances establishes only {\it semidecidability}, not decidability. 

Third, the definition of decidability requires only the existence of a suitable Turing machine; it does not require an explicit construction of such a machine. 
This distinction plays an essential role in our results.

A fundamental example of an undecidable problem is the halting problem, which asks whether a given Turing machine, started on a specified input, eventually halts or runs forever. 
Turing proved that no algorithm can decide this problem for every input.

\bthm{[Undecidability of the halting problem]\lb{t:halting}
The halting problem of Turing machines is undecidable.
}

To understand the difficulty, consider the naive approach of running the given Turing machine and observing its behavior. 
If the machine halts, this procedure confirms its termination after finitely many steps. 
However, if the machine has not halted after a finite number of steps, we cannot determine whether it will eventually halt or continue running forever. 
Thus, this procedure can detect halting but cannot decide non-halting. 
The halting problem is therefore semidecidable, but, as stated in the theorem, not decidable.

\bigskip

We next introduce the notions of provability and unprovability within an axiomatic system.

\bdf{[Provable and unprovable]
A statement $\varphi$ is {\it provable} in an axiomatic theory $T$ if there exists a formal proof of $\varphi$ consisting of a finite sequence of statements, each of which is either an axiom of $T$ or follows from preceding statements according to the rules of inference of $T$. 
If no such proof exists, $\varphi$ is {\it unprovable} in $T$.
}

Note that a statement being unprovable in $T$ does not imply that its negation is provable in $T$.
This point is important in understanding the G\"{o}del's second incompleteness theorem explained later.

In this work, we adopt Zermelo--Fraenkel set theory with the axiom of choice (ZFC), a standard axiomatic foundation of modern mathematics. 
Most of standard mathematics, including algebra, analysis, and topology, can be formalized within ZFC. 
In our setting, frustration-freeness for all system sizes can be formulated as a mathematical statement within ZFC, allowing us to investigate whether this statement is provable in the theory.
We therefore use the provability within ZFC as a formal notion of provability in standard mathematics. 
Throughout this work, unless otherwise specified, provability and unprovability refer to the provability within ZFC.

\bigskip

We next introduce the notions of consistency and inconsistency of an axiomatic theory.
Below, we write $T\vdash\varphi$ if a statement $\varphi$ is provable in a theory $T$, and denote the negation of $\varphi$ by $\neg\varphi$.

\bdf{[Consistent and inconsistent]
A theory $T$ is {\it inconsistent} if there exists a statement $\varphi$ such that both $T\vdash\varphi$ and $T\vdash\neg\varphi$ hold.
Otherwise, $T$ is {\it consistent}.
}

In classical logic, inconsistency has a particularly strong consequence: every statement becomes provable.
This fact is known as the principle of explosion.

\bthm{[Principle of explosion]
If a theory $T$ is inconsistent, then every statement $\psi$ is provable in $T$.
}

\bpf{
Suppose that $T\vdash\varphi$ and $T\vdash\neg\varphi$ for some statement $\varphi$.
For any statement $\psi$, disjunction introduction yields $T\vdash\varphi\lor\psi$.
Combining this with $T\vdash\neg\varphi$ and applying disjunctive syllogism (i.e., $\varphi \lor \psi$ and $\neg \varphi$ means $\psi$), we obtain $T\vdash\psi$.
}

Thus, consistency is essential for an axiomatic theory to distinguish provable statements from those that are
not provable.
ZFC is widely believed to be consistent, and we adopt this assumption in the following discussion of unprovability.

\bigskip

We finally introduce G\"{o}del's second incompleteness theorem, which states that a broad class of consistent axiomatic theories cannot prove their own consistency. 
For a theory $T$, let $\operatorname{Con}(T)$ denote the formal statement that no contradiction can be derived from the axioms of $T$. 
A theory is called {\it recursively enumerable} if its axioms can be enumerated by a Turing machine. 
A standard form of the theorem is stated as follows.

\bthm{[G\"{o}del's second incompleteness theorem]\lb{t:Godel}
Let $T$ be a consistent, recursively enumerable extension of Peano arithmetic (PA). Then $T$ cannot prove its own consistency, i.e.,
\eq{
T\nvdash\operatorname{Con}(T),
}
where $\operatorname{Con}(T)$ is the standard formalization of the consistency of $T$.
}

This theorem also applies to ZFC, whose axioms are recursively enumerable and which is sufficiently expressive to formalize Peano arithmetic. 
Consequently, assuming that ZFC is consistent, we have
\eq{
\mathrm{ZFC}\nvdash\operatorname{Con}(\mathrm{ZFC}).
}
For a detailed proof of the second incompleteness theorem and its mathematical background, see Ref.~\onlinecite{Smibook}.


\subsubsection{Wang tile}\lb{s:Wang}

We here introduce the Wang tiling problem in a slightly generalized form. 
Consider a finite set of square tiles $A^1,\ldots,A^n$, each of which can be placed at any site of the two-dimensional square lattice without rotation or reflection. 
We specify the allowed pairs of neighboring tiles for the horizontal and vertical directions separately. 
In the standard formulation of Wang tiling, each edge of a tile has a color, and two adjacent tiles are compatible if the shared edge by them have matching colors. 
Our formulation allows more general adjacency rules by specifying all the allowed pairs directly. 
The Wang tiling problem asks whether the entire infinite plane can be covered by these tiles with satisfying all local adjacency constraints.

A fundamental result in computability theory is that Wang tiles can simulate the computation of any Turing machine. 
More precisely, for any Turing machine $M$, one can effectively construct a finite set of tiles which tiles the infinite plane if and only if the machine $M$ starting on a blank tape does not halt. 
The key idea to construct such tiles is to encode the computation dynamics by $M$ into arbitrarily large regions generated by a hierarchical tiling structure. 
In Robinson's construction, nested squares appear at arbitrarily large scales. 
Although smaller squares obstruct the interiors of larger ones, unobstructed rows and columns form an effective square grid on which successive rows represent successive configurations of the Turing machine. 
Each such region, called a {\it board}, simulates the machine from a prescribed initial configuration, where a larger board accommodates longer computation. 
Since arbitrarily large boards must occur, a halting machine cannot have a valid tiling of the infinite plane. 
This establishes the undecidability of the Wang tiling problem.

We can further relate the size of the tape space required by a halting Turing machine to the size of finite regions that admit valid tilings. 
Following the boundary treatment in Robinson's construction~\cite{Rob71}, we allow the simulated head to be absorbed when it attempts to leave the available tape interval. 
After the head is absorbed, the tape symbols are propagated unchanged, which does not introduce any forbidden local configuration. 
Thus, the tiling rules impose an obstruction only when the simulated machine reaches its halting state before encountering a boundary. 
Let $\ell$ denote the length of the tape interval visited by a Turing machine before halting. 
Any computation board with fewer than $\ell$ effective tape cells can then be tiled consistently, since the simulated computation must reach a boundary before halting. 
By exploiting the hierarchical structure of Robinson's construction, this property can be used to construct valid tilings of finite squares with side lengths at least $\ell$. 
In contrast, the entire infinite plane cannot be tiled because arbitrarily large boards eventually accommodate the machine's complete halting computation. 
This quantitative connection between tape usage and finite tiling size will be used in \sref{BB}.

We finally discuss the number of tile types required to simulate a Turing machine. 
Let $m$ and $q$ denote the numbers of internal states and tape symbols, respectively. 
In the basic construction of computational Wang tiles, one introduces tape tiles, action tiles, and merging tiles, as described in Robinson's original work~\cite{Rob71} and in Ref.~\onlinecite{Karnote}. 
These three require at most $q$, $mq$, and $2mq$ tile types, respectively, and thus we have at most $(3m+1)q$ computational tiles. 
For a fixed tape alphabet, this number grows linearly with the number of internal states.

The linear scaling also holds for the full construction of the undecidable plane-tiling problem, where the computation is embedded into a hierarchical tiling structure. 
As explicitly noted in Ref.~\onlinecite{DL25}, the number of Wang tile types required in the classical Turing-machine construction scales as $O(mq)$. 
Consequently, for a fixed binary tape alphabet, there exist constants $a,b$ independent of the simulated machine such that the total number of tile types denoted by $N_{\mathrm{tile}}$ satisfies
\eq{
N_{\mathrm{tile}}(m)\leq am+b.
}
This bound directly relates the number of internal states of the Turing machine to the local Hilbert space dimension of the corresponding frustration-free Hamiltonian.


\subsection{Construction of a frustration-free system whose frustration-freeness is unprovable}
\subsubsection{Statement}

Now we shall present our precise statement of unprovability of frustration-freeness.
We say that the frustration-freeness of Hamiltonian $h$ is provable by a theory $T$ if there exists a formal proof with a finite number of symbols satisfying axioms and rules of inference of $T$.
If such a proof does not exist, we say that the frustration-freeness of Hamiltonian $h$ is unprovable by $T$.

Our precise statement of the unprovability is as follows.

\bthm{[Same as \tref{unprovable}]\lb{t:unprovable-s}
Assume that ZFC is consistent.
Then, there exist local Hamiltonians $h^{(1)}$ and $h^{(2)}$ on a two dimensional lattice with the open boundary condition such that (1) this system is frustration-free, and (2) ZFC cannot prove that this system is frustration-free.
}

Accepting the provability within ZFC as a formal notion of provability in standard mathematics and assuming the consistency of our standard mathematics, this theorem claims an astonishing frustration-free system whose frustration-freeness is unprovable by a standard mathematical approach.
We note that the construction of the local Hamiltonians is explicit, and in principle we can write down it within ZFC (though it is extremely complicated and thus we do not write down it in this paper).

\subsubsection{Unprovable frustration-freeness by ZFC}

Below, we shall prove \tref{unprovable-s}.
To this end, we employ the following proposition claiming the presence of a Turing machine which examines the inconsistency of ZFC.
Roughly speaking, this Turing machine writes down all possible proof trees and then checks whether this tree proves a contradiction one by one.

\bpro{[Turing machine testing the consistency of ZFC (Ref.~\onlinecite{YA16})]\lb{t:TM-ZFC}
There exists a Turing machine which halts if and only if ZFC is inconsistent.
}

\bpf{[Proof of \tref{unprovable-s}]
We first show that for any Wang tile there exists a local Hamiltonians $h^{(1)}$ and $h^{(2)}$ on a two dimensional lattice which is frustration-free if and only if the given Wang tile tiles the entire two-dimensional space.
Let $\{ A_i\}$ be a set of tiles in the given Wang tile, which serves as the local Hilbert space of the quantum system.
Let $V_1$ and $V_2$ be sets of horizontal and vertical legal pairs of tiles, respectively.
Then, the desired local Hamiltonians are constructed as 
\eq{
h^{(\alpha)}=P_{V_\alpha^{ \rm c}}=\sum_{A^p- A^q \in V_\alpha^{ \rm c}}\ket{A^pA^q}\bra{A^pA^q}
}
for $\alpha=1,2$, where $P_{V_\alpha^{\rm c}}$ is a projection operator onto the complement of $V_\alpha$ (i.e., set of all illegal pairs).
Since the local Hamiltonian has zero energy if and only if the tiling is legal on this edge, the system is frustration-free if and only if there exists a tiling which tiles the entire space.

We now consider a Wang tile corresponding to the Turing machine testing the consistency of ZFC given in \pref{TM-ZFC}.
Since we have assumed that ZFC is consistent, this Turing machine does not halt, and thus this Wang tile tiles the entire two-dimensional space.
This means that the corresponding quantum system is frustration-free, which proves (1).

On the other hand, if ZFC proves the frustration-freeness of this system, this means that ZFC proves the consistency of ZFC by itself, which contradicts G\"{o}del's second incompleteness theorem (\tref{Godel}).
Hence, ZFC cannot prove the frustration-freeness of this system, which completes the proof of (2).
}

\section{Reconciling decidability and uncomputability}
\subsection{Uncomputability of $L(d)$}

As discussed previously, the decidability of frustration-freeness (\tref{cutoff} and \tref{FF-cutoff-gen}) and the existence of a frustration-free Hamiltonian whose frustration-freeness is unprovable within ZFC (\tref{unprovable-s}) may appear difficult to reconcile. 
A key observation is that the existence of a finite cutoff does not imply our ability to compute it. 
In fact, below we establish that no algorithm can determine a valid cutoff $L(d)$ uniformly in $d$.

\bthm{[Uncomputability of the finite cutoff]\lb{t:uncomputable-s}
For the two-dimensional square-lattice problem with open boundary conditions, there exists no computable function $F(d)$ that provides a valid finite cutoff for any local Hilbert space dimension $d$. 
In particular, the smallest valid cutoff $L_{\min}(d)$ is uncomputable and admits no computable upper bound.
}

To prove this theorem, we first show the following undecidability result.

\bthm{[Undecidability for unrestricted local Hilbert space dimension]\lb{t:undecidable-gend}
The frustration-free problem for translation-invariant nearest-neighbor Hamiltonians on two-dimensional square lattices with open boundary conditions is undecidable when the local Hilbert space dimension $d$ is arbitrary.
}

\bpf{[Proof of \tref{undecidable-gend}]
Recall that, for any Turing machine $M$, we can effectively construct a finite set of Wang tiles that tiles the infinite plane if and only if $M$ does not halt. 
As shown in \sref{Wang}, this tile set can be mapped to a pair of local Hamiltonians which is frustration-free for all system sizes if and only if the corresponding Wang tiles can cover the infinite plane. 
Thus, we obtain an effective construction that maps any Turing machine $M$ to a two-dimensional Hamiltonian that is frustration-free if and only if $M$ does not halt.

If the frustration-free problem with unrestricted $d$ were decidable, we could determine whether any given Turing machine halts by applying the corresponding decision algorithm to this Hamiltonian. 
This contradicts the undecidability of the halting problem (\tref{halting}). 
Hence, the frustration-free problem with unrestricted $d$ is undecidable.
}

\bpf{[Proof of \tref{uncomputable-s}]
Suppose, for contradiction, that there exists a computable function $F(d)$ that provides a valid cutoff for all $d$. 
Given a pair of local Hamiltonians $(h^{(1)},h^{(2)})$ with local Hilbert space dimension $d$, we could first compute $F(d)$ and then determine whether the Hamiltonian on the $F(d)\times F(d)$ square lattice is frustration-free.
The latter problem is decidable by finite-dimensional linear algebra. 
Specifically, frustration-freeness is equivalent to the existence of a nonzero vector in the common kernel of all local Hamiltonians.
Since the matrix elements of the input Hamiltonians are algebraic numbers, whether this common kernel remains nontrivial or not can be computed exactly.

By the definition of the cutoff, this procedure would decide whether the given Hamiltonian is frustration-free for all system sizes, for arbitrary input $d$. 
This contradicts \tref{undecidable-gend}. 
Therefore, no computable function can provide a valid cutoff uniformly in $d$.

Furthermore, any upper bound on the smallest cutoff $L_{\min}(d)$ is itself a valid cutoff, owing to the monotonicity of frustration-freeness for open boundary conditions. 
Hence, $L_{\min}(d)$ cannot have a computable upper bound.
}

Remarkably, this undecidability persists even when all local interactions are restricted to commuting diagonal projection operators with matrix elements in $\{0,1\}$. 
Thus, the undecidability arises already in classical constraint systems and does not rely on intrinsically quantum effects.

In summary, we have thus established two contrasting results:
\bi{
\item The frustration-free problem on the two-dimensional lattice with the open boundary condition with fixed local Hilbert dimension $d$ is decidable for any $d$ (\tref{cutoff}).
\item The frustration-free problem on the two-dimensional lattice with the open boundary condition with arbitrary local Hilbert dimension $d$ is undecidable (\tref{undecidable-gend}).
}
These two results are compatible because decidability for each fixed $d$ guarantees only the existence of a corresponding decision algorithm $M_d$, not a uniform procedure for constructing $M_d$ from $d$ (see \sref{review-halting}). 
Indeed, a finite cutoff can be hard-coded into $M_d$ for each fixed $d$, but no algorithm can determine such cutoffs uniformly. 
Otherwise, a uniform decision algorithm for arbitrary $d$ would exist, contradicting \tref{undecidable-gend}. 
This nonuniformity is precisely what allows fixed-dimensional decidability to coexist with undecidability for unrestricted local dimensions.

\subsection{Scaling of $L(d)$}\lb{s:BB}

The cutoff $L(d)$ grows at least as rapidly as the {\it busy-beaver function}, up to a linear rescaling of its argument. 
The busy-beaver function $\mathrm{BB}(n)$ is defined as the maximum number of 1's remaining on the tape by a halting Turing machine with $n$ internal states and a binary tape alphabet, starting from a blank tape. 
It is known that $\mathrm{BB}(n)$ eventually exceeds any computable function $f(n)$, and hence is itself uncomputable. 
We show that there exist constants $a,b$ such that
\eq{
L(an+b)>\mathrm{BB}(n)
}
for all sufficiently large $n$.

To establish this bound, consider an $n$-state Turing machine $M$ that halts with outputting $\mathrm{BB}(n)$ copies of 1's on the tape. 
Let $s(M)$ denote the width of the smallest tape interval containing all cells visited by the head before halting. 
Since every cell containing a 1 at the end of the computation must have been visited by the head, we have $s(M)\geq\mathrm{BB}(n)$. 
As discussed in \sref{Wang}, a Turing machine with $n$ internal states and a fixed tape alphabet can be encoded into a set of Wang tiles whose number is bounded by $an+b$ for some constants $a,b$.

Below we shall show that the corresponding Wang tiles can cover a finite square with side length at least $s(M)$, although they cannot tile the infinite plane.
Recall that in Robinson's construction, the interior of a hierarchical square contains several computation square boards.
An arbitrarily large computation board can be found in a sufficiently large hierarchical square. 
The computation in each board starts from a prescribed initial configuration. 
When the simulated head reaches the boundary of the available tape interval, it is absorbed, and the tape symbols are thereafter propagated unchanged without violating any tiling constraints. 
Therefore, a board with fewer than $s(M)$ effective tape cells admits a valid tiling: the machine either reaches the boundary before halting or its computation is truncated at the upper boundary.

Since a hierarchical square with side length $s(M)$ contains no complete computation board larger than $s(M)$, a set of Wang tiles corresponding to the machine $M$ tiles a $s(M)\times s(M)$ square.
Finally, the number of tile types, and hence the local Hilbert space dimension, is bounded by $an+b$ for suitable constants $a,b$. 
In conclusion, we obtain
\eq{
L(an+b)\geq s(M)\geq\mathrm{BB}(n),
}
which establishes the claimed busy-beaver lower bound.


\subsection{Comparison with the undecidability of the spectral gap}

It is worth comparing our result with the undecidability of the spectral-gap problem established by Cubbitt {\it et al.}~\cite{CGW15}.
Both problems concern translation-invariant, nearest-neighbor Hamiltonians on two-dimensional square lattices with open boundary conditions, whose local interactions have algebraic matrix elements. 
Although  Cubbitt {\it et al.} also allow an on-site interaction, our decidability theorem extends directly to this setting. 
Remarkably, the spectral-gap problem is undecidable for a sufficiently large but fixed local Hilbert space dimension $d$, whereas frustration-freeness is decidable for any fixed $d$. 
Thus, even for the same class of Hamiltonian inputs, these two fundamental properties exhibit sharply different decidability. 
In particular, the undecidability of the spectral-gap problem demonstrates that the input space (local Hamiltonians with fixed $d$) is large enough to implement an undecidable problem.

A key ingredient in the proof of spectral-gap undecidability is quantum phase estimation embedded in a Feynman-Kitaev type computational history-state Hamiltonian. 
This technique allows us to encode arbitrarily long binary inputs in the phases of local interaction coefficients with keeping the local Hilbert space dimension fixed. 
Our decidability theorem implies that no analogous construction can encode an undecidable problem into frustration-freeness with keeping $d$ fixed. 
Note that this does not exclude quantum phase estimation itself from frustration-free Hamiltonians, but rather excludes its use to establish such an undecidable correspondence. 
Hilbert's basis theorem guarantees that infinitely many system-size constraints reduce to finitely many. 
In contrast, the spectral-gap problem concerns the asymptotic behavior of excitation energies, to which this finite-characterization argument does not apply.


\subsection{Comparison with other related results in computer science}

\begin{table}[t]
\centering
  \begin{tabular}{|l|c|c|c|}\hline
problem &  decidability & origin & cutoff \\ \hline \hline
polynomial automata~\cite{Ben17} &uniformly decidable & Noetherianity & input-dependent cutoff \\ \hline
HDT0L sequence equivalence~\cite{Hon00}  & uniformly decidable & Noetherianity & input-dependent cutoff  \\ \hline
polynomial grammer~\cite{Sch21}  & uniformly decidable & Noetherianity & input-dependent cutoff  \\ \hline
Petri net reachability~\cite{May84}  & uniformly decidable & close to Noetherianity & input-dependent cutoff \\ \hline
word problem~\cite{MNS93, Del00} & nonuniformly decidable & effective finiteness of inputs & common cutoff \\ \hline
locally finite rings~\cite{CDT06}  & nonuniformly decidable & effective finiteness of inputs & common cutoff \\ \hline
graph minor obstruction~\cite{CDF97}  & nonuniformly decidable & effective finiteness of inputs & input-dependent cutoff \\ \hline
parameterized model checking~\cite{Ami14}  & nonuniformly decidable & pathological topology & common cutoff \\ \hline
{\bf frustration-freeness (This article)} & {\bf nonuniformly decidable} & {\bf Noetherianity} & {\bf common cutoff} \\ \hline
ideal membership~\cite{Bau74}  &  nonuniformly decidable & Noetherianity & input-dependent cutoff \\ \hline
spectral gap~\cite{CGW15}  & undecidable &  &  \\ \hline
   \end{tabular}
\caption{A table of decidable and undecidable problems and their origins.
The frustration-free problem is a scarce example where Noetherianity accompanies an uncomputable single common cutoff even on an ordinal square lattice.}\lb{table1}
\end{table}

The distinctive nature of our decidability result becomes clearer when compared with other decidability results in theoretical computer science. 
For a decision problem with only finitely many possible inputs, decidability follows trivially from the existence of a finite truth table, even without an explicit algorithm for constructing it. 
In contrast, fixing the local Hilbert space dimension $d$ does not make the input space of the frustration-free problem finite: even when restricted to algebraic matrix elements, the possible local Hamiltonians form an infinite set with unbounded description lengths. 
Indeed, as explained above the spectral gap problem accompanies the same input space with a fixed  local Hilbert space dimension $d$ and is undecidable~\cite{CGW15}.
Thus, the decidability of frustration-freeness is a nontrivial consequence of the algebraic structure revealed by Hilbert's basis theorem.

Our approach is closely related to the {\it Hilbert method}, which exploits Hilbert's basis theorem to reduce infinitely many algebraic constraints to finitely many. 
This method has been employed in several decision problems in computer science, including the zeroness problem for polynomial automata~\cite{Ben17} and the HDT0L sequence equivalence problem~\cite{Hon00}. 
In the Hilbert method and related approaches~\cite{Ben17, Sei71, Hon00, FS01, Sch21, Boj19, May84}, additional algebraic structures allow effective decision procedures to be constructed, sometimes using techniques such as Gr\"{o}bner bases. 
Specifically, in the zeroness problem for polynomial automata, we have not only a finite-length increasing ideal sequence
\eq{
I_2\subseteq I_3\subseteq I_4 \cdots  \text{ and there exists }N^* \text{ such that }  I_{N}=I_{N^*} \text{ for all } N\geq N^*
}
but also an additional algebraic structure
\eqa{
I_n=I_{n+1} \text{ means } I_n=I_{n+1}=I_{n+2}=\cdots .
}{ideal-stop}
The latter additional structure makes the zeroness problem for polynomial automata to be uniformly decidable.
Some studies employ good upper bounds with which uniform algorithms solving these decision problems can be constructed~\cite{Sei71, Boj19, Sch21}.
In contrast, our application of Hilbert's basis theorem to the frustration-free problem establishes only the existence of a finite cutoff without providing an effective procedure or a good algebraic structure as \eref{ideal-stop}. 
Indeed, our uncomputability theorem shows that no uniform algorithm can compute a valid cutoff $L(d)$ for arbitrary $d$. 
This demonstrates that the existence of a finite algebraic characterization does not necessarily imply the existence of a uniform decision algorithm.

The frustration-free problem provides nonuniform decidability on an infinitely large input space.
Several problems including word problems on mathematical logic~\cite{MNS93, Del00} and locally finite rings~\cite{CDT06} possess nonuniform decidability on an infinitely large input space, though their infinitely many inputs can be essentially reduced to finitely many inputs and the decidability simply comes from this finiteness.
Another problem in mathematical logic, the parameterized model checking problem on general network topology, is shown to be nonuniformly decidable and have a common cutoff~\cite{Ami14}, while this nonuniformity holds only on a special network with a pathological topology, implying that its nonuniformity arises from its pathological network topology rather than the parameterized model checking process itself.
This shows high contrast to the frustration-free problem, where the nonuniformity of its decidability holds even on a two-dimensional square lattice.
Baur~\cite{Bau74} showed a pioneering decidability result on the individual ideal membership problem  in computational algebra carried by Noetherianity.
However, to derive this result we use the Noetherianity of each ideal, and thus we have only an input-dependent finite cutoff, not a common cutoff independent of the input.
Unlike the individual ideal membership problem, the frustration-free problem has a single common ascending chain of polynomial ideals independent of the input local Hamiltonians, which allows us to obtain a common finite cutoff $L(d)$.
It is interesting that nonuniform decidability with a common finite cutoff on an genuinely infinitely large input space arises from a natural problems in physics.


\end{document}